\documentclass[a4paper,11pt]{article}
\pdfoutput=1
\usepackage{jheppub} 
\usepackage[utf8]{inputenc}
\usepackage{amssymb}
\usepackage{amsmath}
\usepackage{amsfonts}
\usepackage{graphicx}
\usepackage{color}
\usepackage{xspace}
\usepackage[normalem]{ulem}

 \usepackage[section]{placeins}
\usepackage{afterpage}

\usepackage{float}
\usepackage{slashed}
\usepackage{multirow,rotating}

\def\gsim{\raise0.3ex\hbox{$\;>$\kern-0.75em\raise-1.1ex\hbox{$\sim\;$}}}
\def\lsim{\raise0.3ex\hbox{$\;<$\kern-0.75em\raise-1.1ex\hbox{$\sim\;$}}}

\newcommand{\ba}[1]{\begin{eqnarray} \label{(#1)}}
\newcommand{\ea}{\end{eqnarray}}

\newcommand{\lam}{\lambda}

\definecolor{tobycolour}{rgb}{.6,.0,.4}

\definecolor{dcolour}{rgb}{.5, .5, .5}

\newcommand{\AddrBonn}{%
Bethe Center for Theoretical Physics \& Physikalisches Institut der 
Universit\"at Bonn,\\ Nu{\ss}allee 12, 
 53115 Bonn, Germany}

\newcommand{\AddrDESY}{%
  Deutsches Elektronen-Synchrotron DESY,\\ Notkestraße 85, 22607 Hamburg, Germany}

\newcommand{\AddrAHEP}{
  {\it AHEP Group, Instituto de F\'{\i}sica Corpuscular --
    CSIC/Universitat de Val{\`e}ncia \\
    Edificio de Institutos de Paterna, Apartado 22085,
  E--46071 Val{\`e}ncia, Spain}}

\def\gsim{\raise0.3ex\hbox{$\;>$\kern-0.75em\raise-1.1ex\hbox{$\sim\;$}}}
\def\lsim{\raise0.3ex\hbox{$\;<$\kern-0.75em\raise-1.1ex\hbox{$\sim\;$}}}

\begin{document}

\preprint{\parbox[t]{3.3cm}{BONN-TH-2018-14 \\ DESY 18-197  \\ IFIC/18-39 }}

\title{Long-Lived Fermions at AL3X}

%\author{Daniel Dercks}
%\email{daniel.dercks@desy.de}
%\affiliation{II. Institut f\"ur Theoretische Physik, Universit\"at Hamburg,\\ Luruper Chaussee 149, 22761 Hamburg, Germany}

%\author{Herbi K. Dreiner}
%\email{dreiner@uni--bonn.de}
%\affiliation{Physikalisches Institut der Universit\"at Bonn, Bethe Center for Theoretical Physics, \\ Nu{\ss}allee 12, 53115 Bonn, Germany}

%\author{Zeren Simon Wang}
%\email{wzeren@physik.uni-bonn.de}
%\affiliation{Physikalisches Institut der Universit\"at Bonn, Bethe Center for Theoretical Physics, \\ Nu{\ss}allee 12, 53115 Bonn, Germany}

\author[a]{Daniel Dercks} \emailAdd{daniel.dercks@desy.de}\affiliation[a]{\AddrDESY}

\author[b]{\!\!, Herbert K. Dreiner} \emailAdd{dreiner@uni-bonn.de}\affiliation[b]{\AddrBonn}

%\author[c]{\!\!, Juan Carlos Helo} \emailAdd{jchelo@userena.cl}\affiliation[c]{\AddrUFSM}

\author[c]{\!\!, Martin Hirsch} \emailAdd{mahirsch@ific.uv.es}\affiliation[c]{\AddrAHEP}

\author[b]{and Zeren Simon Wang}\emailAdd{wzeren@physik.uni-bonn.de}%\affiliation[b]%{\AddrBonn}

%%%%%%%%%%%%%%%%%%%%%%%%%%%%%%%%%%%%%%%%%%%%%%%%%%%%%%%%%%%%%%%%%%%%%%
\abstract{Recently Gligorov \textit{et al.} \cite{Gligorov:2018vkc} proposed to build a cylindrical detector named 
`\texttt{AL3X}' close to the  \texttt{ALICE} experiment at interaction point (IP) 2 of the LHC, aiming for discovery of 
long-lived particles (LLPs) during Run 5 of the HL-LHC. We investigate the potential sensitivity reach of this detector in 
the parameter space of different new-physics models with long-lived fermions namely heavy 
neutral leptons (HNLs) and light supersymmetric neutralinos, which have both not previously been studied 
in this context. Our results show that the \texttt{AL3X} reach can be complementary or superior to that of other proposed 
detectors such as \texttt{CODEX-b}, \texttt{FASER}, \texttt{MATHUSLA} and \texttt{SHiP}. }
%%%%%%%%%%%%%%%%%%%%%%%%%%%%%%%%%%%%%%%%%%%%%%%%%%%%%%%%%%%%%%%%%%%%%%

%\arxivnumber{}
%\pacs{14.60.Pq, 12.60.Jv, 14.80.Cp}

\vskip10mm

\maketitle
\flushbottom
%%%%%%%%%%%%%%%%%%%%%%%%%%%%%%%%%%%%%%%%%%%%%%%%%%%%%%%%%%%%%%%%%%%%%%
%\tableofcontents
%
%%%%%%%%%%%%%%%%%%%%%%%%%%%%%%%%%%%%%%%%%%%%%%%%%%%%%%%%%%%%%%%%%%%%%%
%%%%%%%%%%%%%%%%%%%%%%%%%%%%%%%%%%%%%%%%%%%%%%%%%%%%%%%%%%%%%%%%%%%%%%
%\tableofcontents
%
%%%%%%%%%%%%%%%%%%%%%%%%%%%%%%%%%%%%%%%%%%%%%%%%%%%%%%%%%%%%%%%%%%%%%%

\section{Introduction}

There has recently been an increased interest in neutral long-lived
particles (LLPs). They arise naturally in various models of dark
matter or baryogenesis, for example. For reviews and further models
see Refs.~\cite{Essig:2013lka,Alekhin:2015byh}.  Surprisingly, even
though not designed for this purpose, the LHC detectors \texttt{ATLAS}
and \texttt{CMS}, have a relevant sensitivity to LLPs
\cite{Chatrchyan:2012jna,ATLAS:2012av,Ilten:2015hya,deVries:2015mfw}. However,
there are significant gaps in sensitivity to these models at the LHC. Thus several new
experiments have recently been proposed, to specifically look for
LLPs. These include the beam dump experiment \texttt{SHiP}
\cite{Alekhin:2015byh} at the SPS at CERN. Further experiments are
\texttt{MATHUSLA} \cite{Curtin:2018mvb}, \texttt{CODEX-b}
\cite{Gligorov:2017nwh} and \texttt{FASER} \cite{Feng:2017uoz}, which
would all be located at various positions with respect to the
interaction points (IPs) of \texttt{ATLAS} or \texttt{CMS}, and would
thus make use of LHC events. They are all shielded from the IPs by
between 25m and 450m of rock. Very recently a further experiment has
been proposed, \texttt{AL3X} \cite{Gligorov:2018vkc}, which would be
located at the \texttt{ALICE} site at the LHC. It differs from
\texttt{MATHUSLA}, \texttt{CODEX-b} and \texttt{FASER} in that the
center of the detector is located only 11.25\,m from the
\texttt{ALICE} IP, and furthermore the detector would have a
magnet. Due to the proximity to the IP the experiment would have a
significantly higher geometric acceptance, even for a comparatively
small detector, than the other three proposed new experiments at the
LHC. Such a small detector could be equipped with dense tracking
instrumentation, which would be further improved by the magnetic
field.

A subclass of interesting LLPs are heavy neutral fermions (HNFs). In
this paper we shall focus on two examples: (a) heavy neutral leptons
(HNLs). These are often also called sterile neutrinos in the
literature. However we prefer the term HNL, because within the
experimental neutrino oscillation community sterile neutrinos are
usually identified with neutrinos with masses of order ${\cal O}$(eV),
whereas we shall focus on masses between 0.1 and 10 GeV. (b) The
lightest neutralino in supersymmetry (SUSY), which can decay via
R-parity violating (RPV) interactions. Somewhat surprisingly a light
neutralino with a mass between 0.5 and 5 GeV, which we shall consider, 
is still consistent with all observations 
\cite{Choudhury:1999tn,Dreiner:2003wh,Dreiner:2006sb,Dreiner:2009ic}.\footnote{In fact even a massless neutralino 
is consistent with allobservations \cite{Dreiner:2009ic}.}
Recently we have investigated the search sensitivity of \texttt{MATHUSLA},
\texttt{CODEX-b} and \texttt{FASER} for these specific
HNFs \cite{Helo:2018qej,Dercks:2018eua}.

It is the purpose of this paper to directly extend this recent work
and investigate the sensitivity of the proposed \texttt{AL3X} detector
to these HNFs, \textit{i.e.} sterile neutrinos, as well as the
lightest neutralino in supersymmetry, as they were not considered
in the original \texttt{AL3X} paper \cite{Gligorov:2018vkc}. For the
HNLs we shall consider the production at the \texttt{ALICE} IP via 
$D$-and $B$-mesons. For the neutralinos we also consider the
production via the decay of $D$- and $B$-mesons, and in addition the
direct pair production via $Z$-bosons.

The outline of the paper is as follows. In
Sec.~\ref{sec:detector-and-simulation} we discuss the proposed AL3X
detector set-up and define the parameters for our analysis. We
furthermore present the details of our simulation of the long-lived
HNFs. In Sec.~\ref{sect:HNL} we present our results for the
sensitivity of \texttt{AL3X} to long-lived HNLs. In
Sec.~\ref{sec:light-neutralinos-rpv} we present our results for the
sensitivity of \texttt{AL3X} of long-lived light neutralinos. We
consider separately the pair-production via $Z^0$ decays and the
single production via rare heavy meson decays. In
Sec.~\ref{sec:conslusions} we summarize and offer our conclusions.

\section{Simulation and Detector}
\label{sec:detector-and-simulation}

In this section we outline our simulation procedure and introduce the
setup of the proposed detector \texttt{AL3X}.  Throughout this work we
assume zero background events and $100\%$ detector efficiency. See the
discussion in Ref.~\cite{Gligorov:2018vkc}.

\subsection{Simulation Procedure}

In order to obtain the expected number of detectable decay events, we
estimate the total number, $N_M$, of mother particles $M$ produced at
the LHC from existing experimental results. Here $M$ can be a $D$- or
a $B$-meson, or a $Z$-boson. We then calculate the branching
ratio of the various $M$s into the LLP(s), and compute the average decay probability of these LLPs inside the decay chamber of \texttt{AL3X}. We implement this aspect in a manner very similar to the treatment applied in our previous work, Refs.~\cite{Helo:2018qej,Dercks:2018eua}. Note, that for $\text{BR}(Z \rightarrow \tilde \chi_1^0 \tilde \chi_1^0)$ we have only an experimental upper limit. We will assume two different values of $\text{BR}(Z \rightarrow \tilde \chi_1^0 \tilde \chi_1^0)$ in our numerical study for illustration.

Since the rare decays of charm and bottom mesons into HNLs lead to the
strongest sensitivity reach in HNL mass $m_N$ and mixing square
$|V_{\alpha N}|^2 (\alpha=e, \mu)$, defined in Sec.~\ref{sect:HNL}, we
focus on these channels, discarding the complementary contributions
from $W$-, $Z$- and Higgs bosons.\footnote{The latter were, however,
  taken into account in Ref.~\cite{Helo:2018qej}.} Similarly, in the
case when an RPV $LQ\bar D$ coupling induces single production of a
neutralino, we consider only rare decays of $D$- and $B$-mesons, as
well. From results published by the LHCb collaboration
\cite{Aaij:2015bpa,Aaij:2016avz}, we estimate the number of produced
mesons over a hemisphere for an integrated luminosity of
$\mathcal{L}=100/$fb:
\begin{eqnarray}
N_{D^+}\!\! &= &5.27\times 10^{14},\, N_{D_s^+}=1.70\times 10^{14},\, N_{D_0} = 1.00\times 10^{15}, \label{eq:numofds}\\[2.5mm]
N_{B^+} \!\!&= &2.43\times 10^{13},\, N_{B^0}=2.43\times 10^{13},\, N_{B_s^0} = 5.48\times 10^{12},\, \nonumber \\[2.5mm]
N_{B_c^+} \!\!&= &5.54\times 10^{10}.  
\label{eq:numofbs}
\end{eqnarray}
Besides the LLPs produced from rare meson decays, we furthermore include the case of light neutralinos pair-produced from 
$Z$-boson decays. Experimentally viable light neutralinos must be dominantly bino-like \cite{Choudhury:1999tn,Dreiner:2003wh}, 
with only a small higgsino component. It is the latter, which couples to the $Z$-boson. However, given the large cross section for $Z$-boson 
production at the LHC, we may still obtain good sensitivity reach in $LQ\bar D$ couplings up to a neutralino mass roughly 
half of the $Z$-boson mass. \texttt{ATLAS} published the experimentally measured cross section of $Z\rightarrow \ell^+\ell
^-\, (\ell=e, \mu)$ in $pp$ collisions at $\sqrt{s}=13$ TeV  \cite{Aad:2016naf}. With the BR$(Z\rightarrow \ell^+ \ell^-)$ given 
by the PDG \cite{Tanabashi:2018oca}, we estimate the number of $Z$-bosons produced to be
\begin{eqnarray}
N_{Z} = 2.94\times 10^9,
\end{eqnarray}
over a hemisphere for $\mathcal{L}=100/$fb.

We write the total number of LLPs produced, $N_{\text{LLP}}^{\text{prod}}$, as
\begin{eqnarray}
N_{\text{LLP}}^{\text{prod}}&=&\sum_M N_M\cdot \Gamma(M\rightarrow \text{LLP(s)}+X)\cdot\tau_M\nonumber \\
&=&\sum_M N_M\cdot 
\mathrm{BR}(M\rightarrow \text{LLP(s)}+X),
\end{eqnarray}
where $M$ can be either a $D$- or a $B$-meson, or a $Z$-boson, and with $\tau_M$ denoting its lifetime. To determine 
the average decay probability of the LLPs inside the \texttt{AL3X} ``detectable region''(``d.r.''), $\langle P[\text{LLP}\text{ in d.r.}]
\rangle$, we perform a Monte Carlo (MC) simulation with \texttt{Pythia 8.205} \cite{Sjostrand:2006za,Sjostrand:2014zea}.
We implement the following formula:
\begin{eqnarray}
\langle P[\text{LLP}\text{ in d.r.}]\rangle=\frac{1}{N^{\text{MC}}_{\text{LLP}}}\sum_{i=1}^{N^{\text{MC}}_
{\text{LLP}}}P[(\text{LLP})_i\text{ in d.r.}]\,,
\label{eq:decay_prob}
\end{eqnarray}
where $N^{\text{MC}}_{\text{LLP}}$ is the number of LLPs generated in the MC simulation sample. We generate the $D$- and 
$B$-mesons by making use of the matrix element calculators \texttt{HardQCD:hardccbar} and \texttt{HardQCD:hardbbbar}, 
respectively, of \texttt{Pythia}. In order to extract the kinematics of pair-produced neutralinos from $Z$-boson decays, we resort 
to the ``New-Gauge-Boson Processes" provided by \texttt{Pythia} to generate pure $Z'$-bosons with the same mass as the
Standard-Model (SM) $Z$-boson and let it decay to a pair of new fermion particles. Finally, we calculate the number of 
observed decays of the LLPs in the detector,
\begin{eqnarray}
N^{\text{obs}}_{\text{LLP}}=N_{\text{LLP}}^{\text{prod}}\cdot \langle P[\text{LLP}\text{ in d.r.}]\rangle \cdot 
\text{BR}(\text{LLP}\rightarrow\text{visible only}),
\end{eqnarray}
where we also include $\text{BR}(\text{LLP}\rightarrow\text{visible only})$, the branching ratio of the LLP into only visible states 
such that the event may be reconstructed by \texttt{AL3X}. In Ref. \cite{Gligorov:2018vkc} it was pointed out that the LLP vertex is
required, in order to point back to the IP, and thus be able to reduce the background. 

With \texttt{Pythia} providing the kinematical information of each generated LLP, we can easily derive its velocity $\beta_i$ and 
Lorentz boost factor $\gamma_i$. We calculate the total 
decay width of the HNLs by using the formulas given in Ref. \cite{Atre:2009rg}. As for the decay width of the light neutralinos, we use 
the relevant expressions for neutralino two-body decays given in Ref. \cite{deVries:2015mfw} for a neutralino mass below $\sim 
3.5$~GeV, and take the three-body decay results given by \texttt{SPheno-4.0.3} \cite{Porod:2003um,Porod:2011nf} for 
larger masses. Combining the total decay width $\Gamma_{\text{tot}}(\text{LLP})$ with the $\beta_i$ and $\gamma_i$, we 
express the decay length, $\lam_i$, of a given LLP, $(\text{LLP})_i$, in the laboratory frame:
\begin{eqnarray}
\lambda_i=\beta_i\gamma_i/\Gamma_{\text{tot}}(\text{LLP}),\\
\lambda_i^z=\beta_i^z\gamma_i/\Gamma_{\text{tot}}(\text{LLP}),
\end{eqnarray}
where $\lambda_i^z$ is the $z$-component of $\lambda_i$ along the beam axis. The decay length is required in order to calculate 
the decay probability $P[(\text{LLP})_i\text{ in d.r.}]$. 

\subsection{The \texttt{AL3X} Detector}

\begin{figure}[t]
\centering
 \includegraphics[width=0.45\textwidth]{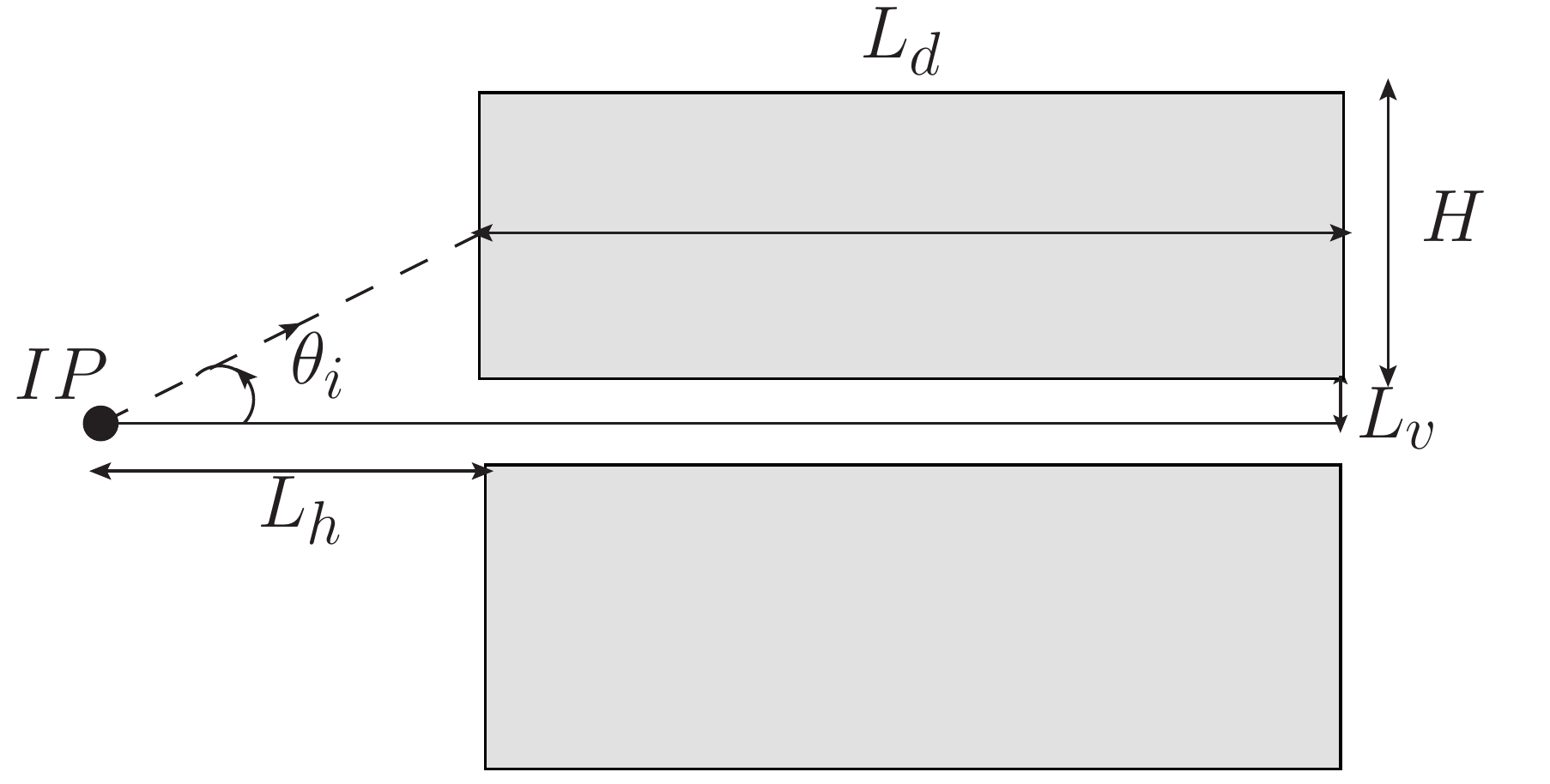}
\caption{Side-view sketch of the \texttt{AL3X} detector with definition of distances and angles used in text. The detector is
cylindrically symmetric around the beam axis. IP denotes the 
interaction point 2 at the LHC. The dashed line describes an example LLP track, with polar angle $\theta_i$. }
\label{fig:al3x-sketch}
\end{figure}

\texttt{AL3X} \cite{Gligorov:2018vkc} is proposed as an on-axis cylindrical detector situated several meters from IP2 in the \texttt{ALICE}/\texttt{L3} cavern at the LHC. It has a length of $L_d = 12\,$m and an inner/outer radius of 0.85/5 m. In 
virtue of its proximity to the IP, its pseudorapidity coverage of $[0.9,3.7]$ is large relative to other proposed future detectors 
such as \texttt{MATHUSLA} $([0.9,1.8])$ \cite{Curtin:2018mvb} and \texttt{CODEX-b} $([0.2,0.6])$ \cite{Gligorov:2017nwh}, 
and it has a full azimuthal coverage.\footnote{\texttt{FASER} $(\eta\gsim 6.9)$ covers a small angular region in the 
extreme forward direction.} We calculate the probability of each individual LLP decaying inside the detector chamber $P[(
\text{LLP})_i\text{ in d.r.}]$ as:
\begin{align}
&P[(\text{LLP})_i\text{ in d.r.}]=e^{-\frac{L_i}{\lambda_i^z}}( 1-e^{-\frac{L'_i}{\lambda_i^z}})\,,\\
L_i&=\text{min}\bigg(\text{max}\bigg(L_h,\frac{L_v}{\tan{\theta_i}}\bigg),L_h+L_d\bigg)\,,\\
L'_i&=\text{min}\bigg(\text{max}\bigg(L_h,\frac{L_v+H}{\tan{\theta_i}}\bigg),L_h+L_d\bigg)-L_i\,,
\end{align}
where $L_h=5.25\,$m is the horizontal distance from the IP to the near end of the detector, $L_v=0.85$ m and $H=4.15$ m 
are respectively the inner radius and the transverse length of the detector, and $\theta_i$ is the polar angle of $(\text{LLP})_i$
with respect to the beam axis. In Ref.~\cite{Gligorov:2018vkc} the authors employed the benchmark integrated luminosities 
100/fb and 250/fb, so that practical concerns such as moving the IP and beam quality, and constraints from backgrounds may 
be investigated. Here we follow their choice of luminosities. In Fig.~\ref{fig:al3x-sketch} we show a profile sketch of \texttt{AL3X}.

Before we present the sensitivity estimates, we compare the average decay probability $\langle P[\text{LLP}\text{ in 
d.r.}]\rangle$, also called fiducial efficencies $\epsilon_{\text{fid}}$ in Ref.~\cite{Gligorov:2018vkc}, of \texttt{AL3X} and
\texttt{MATHUSLA} for the different benchmark models considered in this study. In order to present our results 
for $\epsilon_{\text{fid}}$ with a linear dependence on $c\tau$, we estimate $\epsilon_{\text{fid}}$ in 
the limit that the decay length $\beta \gamma c\tau$ is much larger than the distance from the IP to the detector. 
In the calculation we use the exact formula.
We take the benchmark LLP mass as 1~GeV for both HNLs and light neutralinos. For 
LLPs of 1 GeV produced 
from charm and bottom meson decays, the typical $\beta\gamma$ value is of order $\mathcal{O}(1)$, see Table II in 
Ref.~\cite{Dercks:2018eua}. Thus we require to evaluate $\epsilon_{\text{fid}}$ at $c\tau=100$ m, and our results in Table~\ref{tab:fiducialefficiencies} are roughly valid only for $c\tau\geq 100$ m. On the other hand, LLPs from $Z$-boson 
decays have a $\beta\gamma$ of order $\mathcal{O}(100)$, so our results of $\epsilon_{\text{fid}}$ in this case
are valid roughly for $c\tau\geq 1$ m. The results are shown in Table~\ref{tab:fiducialefficiencies},\footnote{Our estimates 
of the fiducial efficiencies for the lightest neutralinos pair-produced from $Z$-boson decays are somewhat smaller than 
those given in 
Ref.~\cite{Gligorov:2018vkc}, where Higgs decays were considered. This is because Ref.~\cite{Gligorov:2018vkc} applies 
the approximation that $\beta\gamma c \tau$ is much larger than the distance from the IP to the detector in the calculation,
while we use the exact expression.} where $\epsilon_{\text{fid}}^{\text{HNL-D}}, \epsilon_{\text{fid}}^{\text{HNL-B}}, \epsilon_
{\text{fid}}^{\tilde{\chi}_1^0-D_s^+}, \epsilon_{\text{fid}}^{\tilde{\chi}_1^0-B_0}$ and $\epsilon_{\text{fid}}^{\tilde{\chi}_1^0-Z}$ 
denote respectively the fiducial efficiencies for HNLs produced from $D$- and $B$-meson decays, light 
neutralinos produced from $D_s^+$ and $B_0$ decays, and light neutralinos pair-produced from $Z$-boson decays. In general 
one finds that in the large decay length regime \texttt{AL3X} has slightly larger fiducial efficiencies than \texttt{MATHUSLA} in 
these benchmark scenarios.

\begin{table}[t]
\begin{center}
\begin{tabular}{c|c|c|c|c|c}
\hline
{\small Detector} & {\small $\epsilon_{\text{fid}}^{\text{HNL-D}}\cdot c\tau/m$} &{\small $\epsilon_{\text{fid}}^{\text{HNL-B}}
\cdot c\tau /m$ } & {\small$\epsilon_{\text{fid}}^{\tilde{\chi}_1^0-D_s^+}\cdot c\tau /m$} &{\small $\epsilon_{\text{fid}}^
{\tilde{\chi}_1^0-B_0}\cdot c\tau /m$ }&  {\small$\epsilon_{\text{fid}}^{\tilde{\chi}_1^0-Z}\cdot c\tau /m$ }  \\
\hline
\texttt{AL3X} & $4.8\times 10^{-1}$ & $4.6\times 10^{-1}$  & $3.9\times 10^{-1}$ & $2.3\times 10^{-1}$ & $1.6\times 10^{-2}$\\
\texttt{MATHUSLA} & $9.7\times 10^{-2}$ & $1.3\times 10^{-1}$  & $1.1\times 10^{-1}$ & $1.2\times 10^{-1}$ & $8.0\times 10^{-4}$\\
\hline
\end{tabular}
\caption{Summary of fiducial efficiencies of \texttt{AL3X} and \texttt{MATHUSLA} for different models with the benchmark LLP mass 
at 1 GeV.}
\label{tab:fiducialefficiencies}
\end{center}
\end{table}

\section{Heavy Neutral Leptons} 
\label{sect:HNL}

In this section we discuss the prospects of \texttt{AL3X} for detecting heavy
neutral leptons HNLs.  HNLs, $N_j$, have charged (CC) and neutral
current (NC) interactions, suppressed relative to electro-weak
strength via small mixing elements:
\begin{eqnarray}\label{CC-NC}
{\cal L} &=& \frac{g}{\sqrt{2}}\, 
 V_{\alpha N_j}\ \bar \ell_\alpha \gamma^{\mu} P_L N_{j} W^-_{L \mu} 
+\frac{g}{2 \cos\theta_W}\ \sum_{\alpha, i, j}V^{L}_{\alpha i} V_{\alpha N_j}^*  
\overline{N_{j}} \gamma^{\mu} P_L \nu_{i} Z_{\mu},
\end{eqnarray}
where $i=1,2,3$ and $j=1, .., n$, and $\ell_\alpha$, $\alpha=e,\,\mu$,
are the charged leptons of the SM. For kinematic reasons, we restrict
ourselves to the first two generations. $V_{\alpha N_j}$ denotes the
mixing between ordinary neutrinos and the HNLs of mass $m_{N_j}$. The mixing $|V_{\alpha N_j}|$ controls both,
production and decay of the HNLs.

HNLs/sterile neutrinos are mostly motivated by their connection with
the generation of masses for the light, active neutrinos.  In the
standard minimal seesaw picture one simply adds three fermionic
singlets to the SM, together with their Majorana mass terms. Within
this simplest model, one expects that these steriles mix with the
active neutrinos roughly at the order of $V_{\alpha N_j}
\propto\sqrt{m_{\nu}/m_N}$, \textit{i.e.} $|V_{\alpha N_j}|^2 \simeq 5
\times 10^{-11}(\frac{m_{\nu}}{\rm 0.05 eV})(\frac{\rm 1\,GeV}{m_N})$.
However, other model variants, such as the inverse seesaw
\cite{Mohapatra:1986bd}, lead to much larger mixing, despite the
smallness of the observed neutrino masses.   Below, we take $|V_{\alpha N_j}|^2$ as a free parameter in our
calculations.\footnote{See also Ref.~\cite{Dreiner:2008tw} for a
  detailed computation of the see-saw model.}

\begin{figure}[t]
\centering
\includegraphics[scale=0.8]{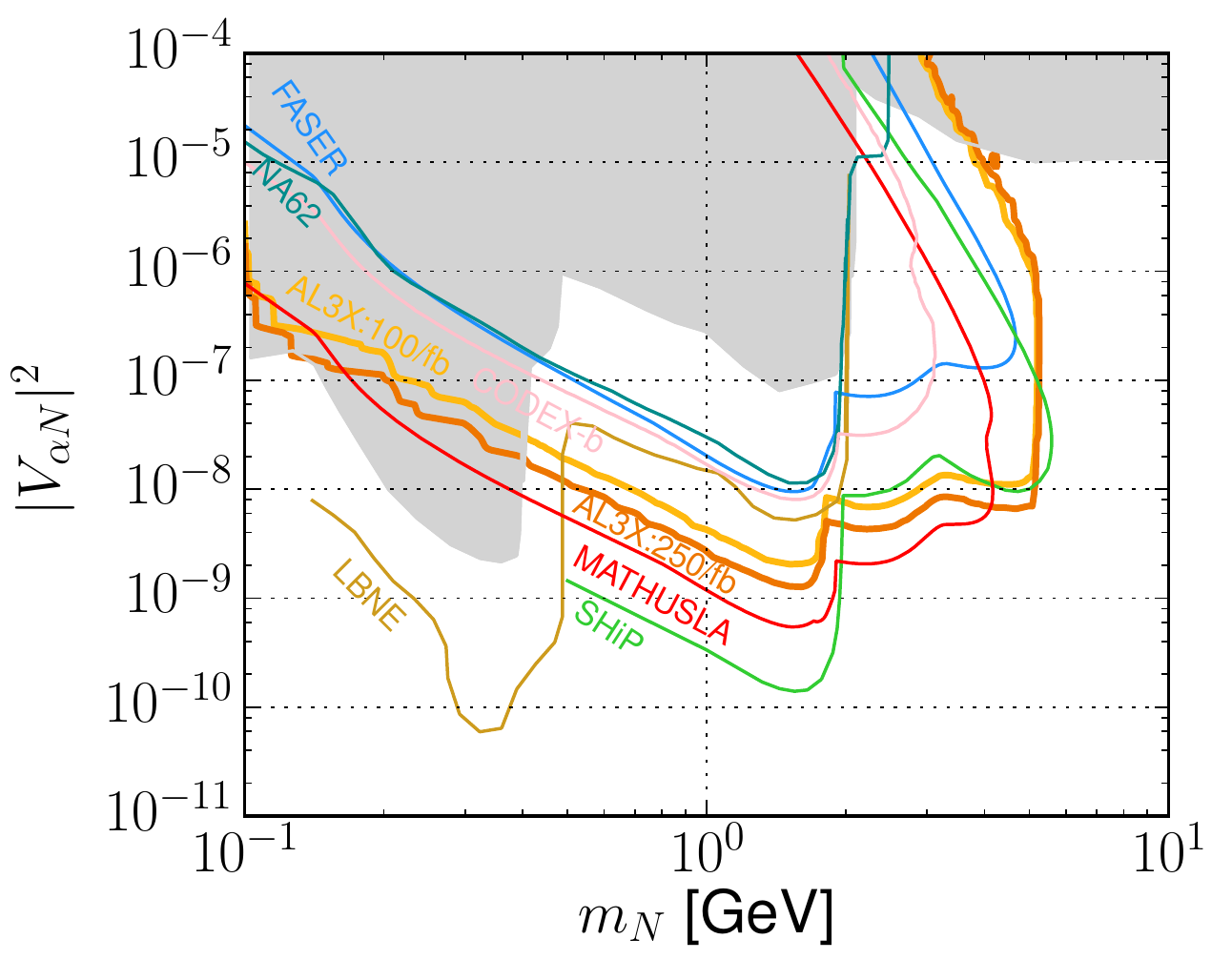}
\caption{Estimates for the sensitivity of different experiments to HNLs in the plane mixing angle squared, $|V_{\alpha N}|^2$, 
versus mass of the HNL, $m_{N}$ [GeV].  The references for the individual curves are given in the text.}
\label{FIG:VsqMN}
\end{figure}

We now turn to the discussion of the results. Fig.~\ref{FIG:VsqMN}
shows sensitivity estimates for \texttt{AL3X} and various other recent
experimental proposals to HNLs. For \texttt{AL3X} we show two curves,
one for $100/$fb and one for $250/$fb, corresponding to the two options
discussed in Ref.~\cite{Gligorov:2018vkc}. The grey area in the
background shows the parameter space currently excluded
according to Ref.~\cite{Deppisch:2015qwa} by  the searches from
\texttt{PS191} \cite{Bernardi:1987ek}, \texttt{JINR}
\cite{Baranov:1992vq}, \texttt{CHARM} \cite{Bergsma:1985is}, and
\texttt{DELPHI} \cite{Abreu:1996pa}. Sensitivities for HNLs for
\texttt{CODEX-b} (300/fb) \cite{Gligorov:2017nwh},
\texttt{FASER} (3/ab) \cite{Feng:2017uoz} and
\texttt{MATHUSLA} (3/ab) \cite{Chou:2016lxi} have
been calculated in Ref.~\cite{Helo:2018qej}. While we use
Ref.~\cite{Helo:2018qej} in this plot, we note that these estimates
agree quite well with other calculations for the same experiments in
Ref.~\cite{Curtin:2018mvb} (for \texttt{MATHUSLA}) and
Ref.~\cite{Kling:2018wct} (\texttt{FASER}). The line for \texttt{LBNE}
is taken from Ref.~\cite{Adams:2013qkq}, \texttt{SHiP}
($2\times 10^{20}$ protons on target) from
Ref.~\cite{Bondarenko:2018ptm}, while the final sensitivity of
\texttt{NA62} was recently estimated in Ref.~\cite{Drewes:2018gkc}.

Fig. (\ref{FIG:VsqMN}) shows that \texttt{AL3X} is quite competitive for the search of HNLs, with a sensitivity better than 
\texttt{FASER}, \texttt{CODEX-b} or \texttt{NA62}, even for only $100$/fb of statistics. In the mass range above $m_N 
\sim 2\,$GeV, \texttt{AL3X} has a sensitivity that is better than the estimate for \texttt{SHiP} \cite{Bondarenko:2018ptm}, 
and only slightly worse than \texttt{MATHUSLA}. Below $m_N \sim 2\,$GeV, \texttt{SHiP} gives the best sensitivity, with 
\texttt{AL3X}$@250$/fb only roughly a factor ($2-3$) less sensitive than \texttt{MATHUSLA} in that mass range. 
Note, however, that the estimate for \texttt{MATHUSLA} is based on $3$/ab of statistics.

\section{Light Neutralinos Decaying via R-Parity Violation}
\label{sec:light-neutralinos-rpv}

We continue with a discussion of the expected sensitivity of \texttt{AL3X} to a light long-lived neutralino in RPV SUSY
\cite{Dreiner:1997uz,Barbier:2004ez}. Supersymmetric theories are an interesting extension to the SM 
\cite{Nilles:1983ge,Martin:1997ns} . In supersymmetry, the fermionic partners of the neutral gauge bosons and the 
neutral CP-even scalar Higgs fields mix to form four mass eigenstates called neutralinos, and denoted $\tilde \chi_i^0$. 
The lightest of these,  $\tilde \chi_1^0$, is typically the lightest particle of the supersymmetric spectrum (LSP).

Rules of constructing gauge-, Lorentz- and SUSY-invariant Lagrangians reproduce the known interactions of the SM, and however additionally predict the following operators in the superpotential \cite{Weinberg:1981wj}
\begin{align}
W_{\text{RPV}}= &\kappa_i L_i H_u + \lam_{ijk}L_i L_j E^c_k + \lam'_{ijk}L_i Q_j D^c_k + 
\lam''_{ijk} U_i^c D_j^c D_k^c\,.
\end{align}
The presence of the dimensionful parameters $\kappa_i$, and/or the Yukawa couplings $\lam_{ijk}$ and/or $\lam^\prime_{ijk}$ 
leads to lepton-number violation, whilst a non-vanishing $\lam^{\prime \prime}_{ijk}$ violates baryon-number. In our study we only 
discuss the phenomenology of non-vanishing $\lam^\prime$. This choice conserves baryon number and hence does not 
lead to unobserved decays of the proton. See also 
Refs.~\cite{Dreiner:2012ae,Dreiner:2005rd,Dreiner:2006xw,deCampos:2007bn,Desch:2010gi,Dercks:2017lfq} on the motivation 
for this choice of couplings and on the changes in phenomenology due to RPV. The $LQ\bar D$ operators predict, among others, 
the following effective operators between the neutralino, and the SM fermions $u, d, \ell$ and $\nu$:
 \begin{align}
\mathcal{L} \supset G^{S,\nu}_{iab} (\overline{\widetilde{\chi}^0} P_L  \nu_i)  (\overline{d_b} P_L  d_a) + G^{S,\ell}_{iab}
(\overline{\widetilde{\chi}^0} P_L  \ell_i) (\overline{d_b} P_L  u_a) + \text{h.c.} \,.\label{eq:interactions4}
\end{align}
The effective couplings $G$ depend on several masses of the scalar supersymmetric partners $\tilde f$ of the SM fermions, the 
mixing within the neutralino sector and are linear in $\lam^\prime$. The exact formulae can be found in Ref.~\cite{deVries:2015mfw}. 
If all $\tilde f$ are mass degenerate $G$ can be written as $\mathcal{O}(1) \times \lam^\prime_{iab}/m^2_{\tilde f}$. Bounds on 
various combinations of $\lam^\prime_{iab}$ and $m_{\tilde f}$ can be set from searches for exotic decays in the meson sector, 
see \textit{e.g.}\ Ref.~\cite{Allanach:1999ic,Kao:2009fg,Dreiner:2009er,Domingo:2018qfg}. In the special case of mass degenerate 
$\tilde f$ they can be compared to our sensitivity curves as we show below. 

The absence of any R-parity violating terms predicts a stable $\tilde \chi_0$. In contrast, the terms in Eq.~(\ref{eq:interactions4}) 
directly imply a long-lived particle, which eventually decays into SM fermions.

\subsection{Pair Production of $\tilde{\chi}_1^0$ from $Z$-Boson Decays}

There are various possibilities to produce neutralinos at the LHC. One of these is the decay of on-shell  $Z$ bosons into pairs
of neutralinos if $m_{\tilde{\chi}_1^0}\lsim m_Z/2$.  The corresponding partial decay width $\Gamma(Z\rightarrow \tilde{\chi}_1^0 \tilde{\chi}_1^0)$ 
has been calculated in Ref.~\cite{Bartl:1988cn} and is proportional to (\textit{cf.} Eq.~(K.2.5) in Ref.~\cite{Dreiner:2008tw})
\begin{equation}
(|N_{13}|^2 - |N_{14}|^2)^2\,, 
\label{eq:higgsino}
\end{equation}
where $N_{13},\,N_{14}$ are the two neutral CP-even higgsino admixtures of the lightest neutralino. A light neutralino is dominantly bino but can have a substantial higgsino admixture \cite{Choudhury:1999tn,Dreiner:2009ic,Helo:2018qej}. However, even in that
case, we see from Eq.~(\ref{eq:higgsino}), that there can be a cancellation leading to a vanishing $Z\to \tilde \chi_1^0\tilde \chi_1^0$
branching ratio. Therefore, the invisible $Z$ decay width can in principle be arbitrarily small. Rather than scan over the 
supersymmetric parameter space, we use $\text{BR}(Z \rightarrow \tilde \chi_1^0 \tilde \chi_1^0)$ as a free parameter, see \textit{e.g.}\ 
Ref.~\cite{Bartl:1988cn} for the detailed dependence.

As discussed in the previous work of Ref.~\cite{Dreiner:2009ic,Helo:2018qej}, typical values of the invisible branching ratio
in supersymmetry is around $6 \times 10^{-4}$, while the experimental upper bounds on the invisible $Z$ branching ratio 
require values below $0.1\,$\% at 90\,\%\,CL, according to Ref.~\cite{Patrignani:2016xqp}. For our analysis we therefore choose 
BR$(Z \rightarrow2\tilde\chi_1^0)$ = $10^{-3}$ and $10^{-5}$, as two representative and experimentally viable values for 
this invisible branching ratio.

For this benchmark analysis, we choose $\lam^\prime_{112}$ to be the only non-vanishing RPV operator. In that case, the 
neutralino can decay into $u+\bar{s}+e^- $and $d+\bar{s}+\nu_e$ final states, as well as their respective charge conjugates. We 
use these inclusive final states to calculate the total lifetime of the neutralino, see Ref.~\cite{Helo:2018qej}. However, in practise 
it may only be feasible to detect charged final states with light mesons, $\tilde \chi_1^0 \rightarrow K^\pm e^\mp$. We also 
calculate the partial decay width into this particular final state according to Ref.~\cite{deVries:2015mfw} and multiply with the 
corresponding branching ratio. We determine results for both cases, \textit{i.e.}\ if only the charged meson final state or if all 
hadronic final states can be observed.

\begin{figure}
\centering
\includegraphics[width=0.45\textwidth]{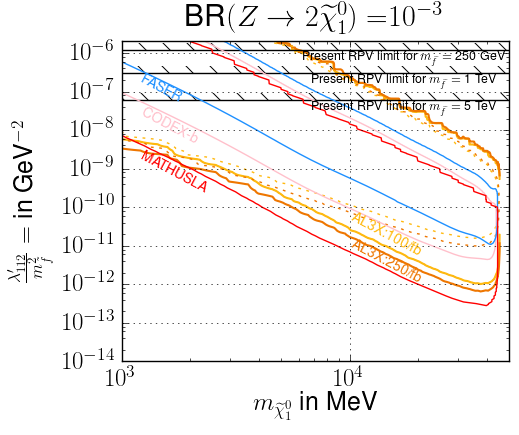}
\;\;  \includegraphics[width=0.45\textwidth]{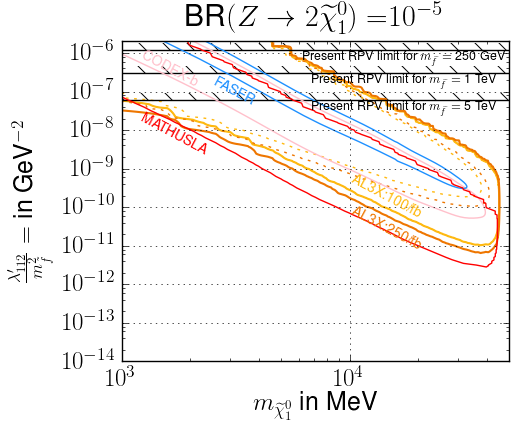}
  \caption{\texttt{AL3X} sensitivity shown in the plane spanned by the neutralino mass and the effective RPV coupling $\lam^\prime
  _{112}/m_{\tilde f}^2$ for two different assumptions of the $Z$ branching ratio to neutralinos. Sensitivity curves denote the 
expected measurement of 3 visible events with an integrated luminosity of 100/fb and 250/fb. Also shown for comparison are the 
sensitivities of other experiments, taken from Ref.~\cite{Helo:2018qej}. Solid lines consider all hadronic final states while dashed 
lines --- only evaluated for \texttt{AL3X} in this work --- only consider the branching ratio into the charged $K^\pm e^\mp$ final 
state for observable neutralino decays. Overlaid current RPV limits on $\lam^\prime_{112}$ are shown for comparison, using 
different assumptions on the degenerate sfermion mass $m_{\tilde f}$. The references are given in the text.}
  \label{fig:zneuneuresults}
\end{figure}

Results for this scenario are shown in Fig.~\ref{fig:zneuneuresults} for both benchmark values of the BR$(Z \rightarrow2\tilde
\chi_1^0)$. We choose the mass of the neutralino as one free parameter, which for kinematic reasons must be smaller than 
$m_Z/2$. As explained above, RPV-induced decays of the neutralino depend on the effective coupling $\lam^
\prime/m_{\tilde f}^2$, which is why we choose this as our second free parameter. Current limits on the RPV operators $L_1 Q_1 
\bar D_2$ are taken from Ref.~\cite{Kao:2009fg} and compared to our results. Note, however, that such a comparison is only 
valid if all sfermions are mass degenerate. See the discussion in Ref.~\cite{deVries:2015mfw}. 

We observe that for invisible branching ratios close to the current PDG limit, \texttt{AL3X} is sensitive to values of $\lam^\prime_{112}/
m_{\tilde f}^2$ down to $10^{-12}$ GeV$^{-2}$, if all hadronic final states can be observed. This strongest sensitivity  is reached for 
neutralino masses near the kinematic threshold, at $m_{\tilde \chi_1^0} \approx 40$~GeV. The sensitivity drops by nearly one order 
of magnitude, if only the charged final state $K^\pm e^\mp$ is taken into account. Note that this search is more sensitive than current 
limits on the $\lam^\prime_{112}$ coupling by several orders of magnitude. The lighter the neutralino, the lower the sensitivity on $\lam
^\prime_{112}/m_{\tilde f}^2$ but even for $\mathcal{O}$(GeV) masses \texttt{AL3X} may be expected to significantly improve on 
current limits. Note that smaller neutralino masses have a reduced accessible final state phase space and hence a reduced difference 
in sensitivity between the conservative ``$K^\pm e^\mp$'' and the optimistic ``all hadronic final states''.

In comparison to the other proposed experiments, \texttt{AL3X} outperforms \texttt{FASER} and \texttt{CODEX-b} over the entire parameter range. It is competitive with \texttt{MATHUSLA} at the low-value range of $\lam'_{112}$ and neutralino masses of a few GeV. For larger masses, \texttt{MATHUSLA} shows the strongest sensitivity for small RPV operators.
%\sout{but is significantly more sensitive at the higher range of $\lam'_{112}$.} \dacomment{moved to discussion in next paragraph}

Note that too large values of $\lam^\prime$ render the neutralino too short-lived to reach the detector which leads to upper 
bounds on the sensitivity to $\lam^\prime$ for all LLP experiments. In comparison to \texttt{FASER}, \texttt{MATHUSLA} and 
\texttt{CODEX-b}, \texttt{AL3X} covers the largest region of parameter space here, due to its proximity to IP2. However, 
for masses above 10~GeV, sizable RPV couplings may still evade both current detection limits and even the limits from 
\texttt{AL3X@250/fb}.

\subsection{Single Production of $\tilde{\chi}_1^0$ from Rare $D$- and $B$-Meson Decays}

\begin{table}
\begin{center}
\begin{tabular}{r||l l}
 & Scenario 1 & Scenario 2 \\
\hline
$\lam^\prime_{\text{prod}}$ for production & $\lam^\prime_{122}$ & $\lam^\prime_{131}$ \\
$\lam^\prime_{\text{dec}}$ for decay & $\lam^\prime_{112}$ & $\lam^\prime_{112}$ \\
produced meson(s) & $D_s$ & $B^0, \bar{B}^0$ \\
visible final state(s) & $K^{\pm} e^\mp, K^{*\pm} e^\mp$ & $K^{\pm} e^\mp, K^{*\pm} e^\mp$ \\
invisible final state(s) via $\lam^\prime_{\text{prod}}$ & $(\eta, \eta^\prime, \phi) + (\nu_e, \bar{\nu}_e)$ & none \\
invisible final state(s) via $\lam^\prime_{\text{dec}}$ & $(K^0_L,K^0_S, K^*) + (\nu_e, \bar{\nu}_e)$ & $(K^0_L,K^0_S, K^*) + (\nu, \bar{\nu})$ \\
\hline
\end{tabular}
\end{center}
\caption{Features of the R-parity violating benchmark scenarios studied in this section.}
\label{tab:rpvbenchmarks}
\end{table}

The effective operators in Eq.~(\ref{eq:interactions4}) mediate interactions between the neutralino, the SM leptons and SM 
mesons. Depending on the masses, a single operator can either predict the decay of a SM meson $M$ into $\tilde \chi_1^0
+\ell$ or predict the decay of a neutralino into $\ell + M$ with small width and hence long lifetime. Here $\ell$ denotes 
either a charged or neutral lepton. In this study we assume that in any given model there are two non-vanishing 
couplings: $\lam^\prime_{\text{prod}}$, $\lam^\prime_{\text{dec}}$, at a time. These are respectively responsible for the 
production of neutralinos via the decay of a heavier meson, and for the decay of neutralinos into a lighter mesons. For simplicity,
we here only consider low-energy models and do not take the effect of generating additional couplings via renormalization
group equations into account \cite{Agashe:1995qm,Allanach:1999mh,Allanach:2003eb}.

In this study, we exemplarily choose two benchmark scenarios with different choices for the non-vanishing $\lam^\prime$, 
summarized in Table~\ref{tab:rpvbenchmarks}. We choose these two scenarios because they are representative for a class of $LQ\bar D$ couplings combinations. For more details, see the discussion below.
For each scenario we have a different initial meson flavor, which produces the 
neutralinos via decays. This is important as these differ in their LHC production yields, see Eqs.~(\ref{eq:numofds}) and 
(\ref{eq:numofbs}), as well as the different final states the neutralinos decay into. Due to the simultaneous presence of several 
operators from one $\lam^\prime$ coupling, see Eq.~(\ref{eq:interactions4}), we often expect both charged and neutral final 
states. We call the former ``visible'', as only those can be experimentally measured by an LLP experiment. We need to consider 
all possible final states for the total lifetime of the lightest neutralino, $\tau_{\tilde \chi}$, but multiply the final number with the 
``visible branching ratio'', \textit{i.e.} the fraction of decays into a charged final state. More details on these benchmarks, 
including formulae for the respective decay widths and branching ratios can be found in Ref.~\cite{deVries:2015mfw}.

\begin{figure}
\centering
  \includegraphics[width=0.45\textwidth]{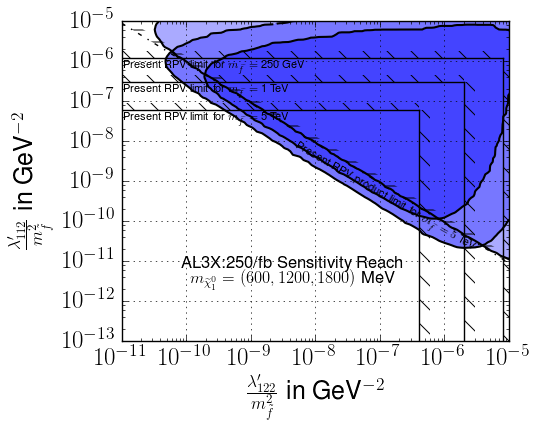}
  \includegraphics[width=0.45\textwidth]{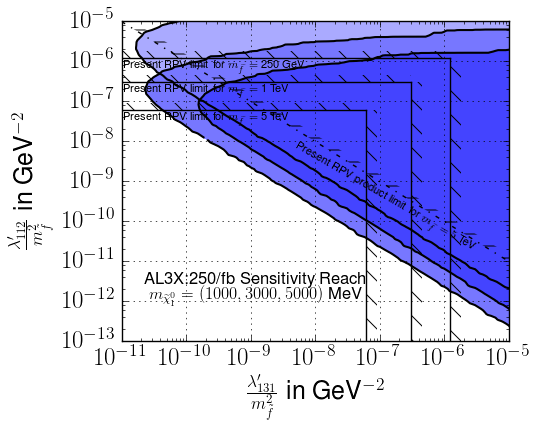}
  \caption{\texttt{AL3X} sensitivity shown in the plane spanned by the two free parameters $\lam^\prime_{\text{prod}}$ and 
  $\lam^\prime_{\text{dec}}$ as respectively defined in Table~\ref{tab:rpvbenchmarks}. Sensitivity curves denote the expected 
measurement of 3 visible events with an integrated luminosity of 250/fb for three different choices for the neutralino mass. The 
first/second/third value denoted in the plot respectively corresponds to the light blue/medium blue/dark blue region, respectively. 
Overlaid RPV-limits are shown for different choices of the sfermion mass $m_{\tilde f}$. }
  \label{fig:rpv1}
\end{figure}

In Fig.~\ref{fig:rpv1} we show model-dependent sensitivity curves of the \texttt{AL3X} detector with an integrated luminosity of 
250/fb. As explained above, the responsible effective operators scale with $\lam^\prime/m_{\tilde f}^2$, which is why we 
choose this ratio of parameters for our figure axes. Since the mass of the long-lived neutralino affects the final state kinematics 
and the accessible phase space, we choose three representative mass values for each scenario: $m_{\tilde \chi_1^0}$ = 600, 
1200 and 1800~MeV for scenario 1 and 1000, 3000 and 5000 MeV for scenario 2. Due to the required decay chain, 
scenario 1 is restricted to the mass range $m_K \leq m_{\tilde \chi_1^0} \leq m_{D_s}$ while scenario 2 requires $m_K \leq 
m_{\tilde \chi_1^0} \leq m_{B}$. For each figure, we also show the corresponding existing bound on the respective RPV operator 
taken from Ref.~\cite{Kao:2009fg}. As mentioned, these can only be compared in the special case of mass degenerate sfermions.

For scenario 1 (2), \texttt{AL3X} is sensitive to values of $\lam^\prime_{\text{prod}}$ down to $\lam^\prime_{122}/m_{\tilde f}^2$ 
($\lam^\prime_{131}/m_{\tilde f}^2) = 3 \times 10^{-11}$ ($2\times 10^{-11}$) GeV$^{-2}$, in case of relatively light neutralino masses, 
\textit{i.e.}\ close to the lower mass threshold. Heavier masses can weaken this bound by up to an order of magnitude. The general 
bounds in the parameter planes then depend on the combination of both $\lam^\prime_{\text{prod}}$ and $\lam^\prime_
{\text{dec}}$ since both couplings can simultaneously affect the neutralino lifetime and its visible branching ratio.

Note that for scenario 1, there exist both upper and lower bounds on $\lam^\prime_{\text{dec}}/m_{\tilde f}^2$=$\lam'_{112}/m_{\tilde 
f}^2$ whose precise values depend on the neutralino mass. If this coupling is chosen too large, the neutralino lifetime is too short 
and they decay before reaching \texttt{AL3X}. If it is too small, however, many neutralinos will live too long and hence one requires 
a large value for $\lam^\prime_{\text{prod}}$ to produce enough neutralinos to still predict three observed decays. However, too large 
values of $\lam^\prime_{\text{prod}}/m_{\tilde f}^2=\lam^\prime_{122}/m_{\tilde f}^2$ predict a too large branching ratio of the 
neutralino into invisible final states and hence too small values of $\lam^\prime_{112}/m_{\tilde f}^2$ below $1 \times 10^
{-11}\,$GeV$^{-2}$ cannot be probed, regardless of the value of $\lam^\prime_{\text{prod}}/m_{\tilde f}^2$. This effect becomes 
more prominent in the results shown in a different parameter plane below.

For scenario 2, $\lam^\prime_{\text{prod}}$ does not produce any invisible final state from neutralino decays, see 
Table~\ref{tab:rpvbenchmarks}, which is why there is no lower bound on the sensitivity to $\lam^\prime_{\text{dec}}/m_{\tilde 
f}^2 = \lam^\prime_{112}/m_{\tilde f}^2$. For this scenario, \texttt{AL3X} is sensitive as long as the product $\lam^\prime_{112}\lam 
^\prime_{131}/m_{\tilde f}^4$ is larger than $\approx 2 \times 10^{-18}$ GeV$^{-4}$.

\begin{figure}
\centering
  \includegraphics[width=0.45\textwidth]{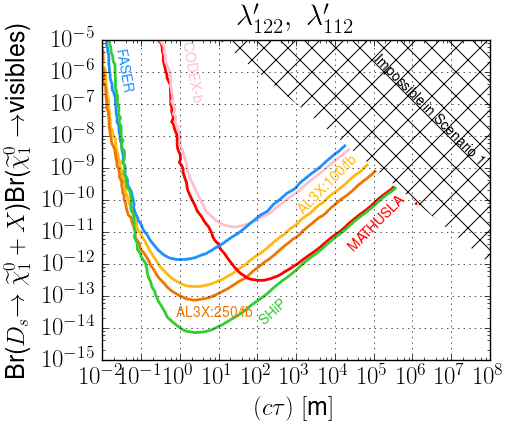}
  \includegraphics[width=0.45\textwidth]{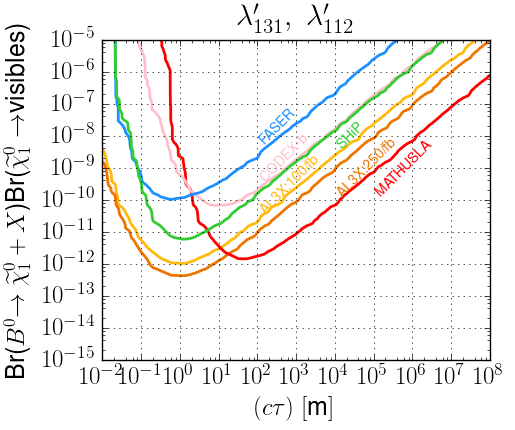}
  \caption{Model-independent sensitivity estimates for different experiments. We show the sensitivity reach as isocurves 
  of 3 events of visible decays. For the axes, we choose the neutralino's unboosted decay length $c \tau$ and the relevant 
  meson branching ratio times the relevant neutralino branching ratio. For scenario 1, regions with large $c \tau$ and large branching ratio are impossible to construct theoretically.}
  \label{fig:rpv2}
\end{figure}

In Fig.~\ref{fig:rpv2}, we show the same results in a different parameter plane, \textit{i.e.}\ the overall branching ratio of the initial 
state mesons into visible final states, $\text{BR}(\text{Meson} \rightarrow \tilde \chi_1^0 + X) \times \text{BR}(\tilde \chi_1^0 
\rightarrow \text{charged final state})$ vs the decay length, $c \tau$, of the neutralino. Here we also overlay results from other 
LLP experiments, \textit{i.e.}\ \texttt{MATHUSLA}, \texttt{CODEX-b} and \texttt{FASER}, as determined in Ref.~\cite{Dercks:2018eua}, 
as well as \texttt{SHiP} from Ref.~\cite{deVries:2015mfw}. We also compare the impact of integrated luminosity and show the 
expected sensitivity at \texttt{AL3X} for both 100/fb and 250/fb. Remember that in scenario 1, increasing  $\lam^\prime_{\text{prod}}$ 
simultaneously increases $\text{Br(Meson} \rightarrow \tilde \chi_1^0 + X$) and decreases $c \tau$. For that reason there exists a 
region in this parameter plane which is theoretically impossible and we marked this as the hashed region in the upper 
right corner in the left plot of Fig.~\ref{fig:rpv2}. No such  effect exists for scenario 2.

For scenario 1, visible branching ratios down to $\approx 8 \cdot 10^{-14}$ can be probed with \texttt{AL3X} while for scenario 2 the 
sensitivity is about a factor of 5 weaker. The difference can be explained by considering the difference in meson production rates, see
 Eqs.~(\ref{eq:numofds}) and (\ref{eq:numofbs}), the respective fiducial efficiencies, see Table~\ref{tab:fiducialefficiencies}, and 
 the effect of invisible decays only present in the first scenario, see Table \ref{tab:rpvbenchmarks}. 

 In comparison with the other experiments, we see that \texttt{AL3X} outperforms both \texttt{FASER} and \texttt{CODEX-b} in both 
 scenarios. As scenario 1 relies on the abundant production of $D$-mesons, \texttt{SHiP} which works at a centre-of-mass energy 
 of $\approx 27$~GeV is significantly more sensitive. \texttt{AL3X} can only improve on the expected \texttt{SHiP} bound for scenarios with mean decay paths below 0.1~m, due to the proximity of the detector to IP2.

Comparing to the expected sensitivity curves from \texttt{MATHUSLA}, we observe that \texttt{AL3X} can obtain far stronger results 
for models, with $c \tau$ below roughly $20$~m. Again, this can be explained by the different geometry: whilst \texttt{AL3X} 
is designed with a target-to-detector-distance of about 5~m close to the IP, \texttt{MATHUSLA} is planned as a surface experiment 
with a respective distance of more than 140~m.

\section{Conclusions}
\newcommand{\nicefrac}[2]{#1/#2}
\label{sec:conslusions}
In this work we have investigated the sensitivity of the recently proposed detector \texttt{AL3X} for detecting long-lived fermions 
in the context of heavy neutral leptons (HNLs), also known as singlet neutrinos, and the lightest neutralino of 
supersymmetry. For the HNLs study, we present results where solely the mixing between $\nu_{e/\mu}$ and the HNL, $N$, is 
non-vanishing. For the neutralino case, we consider two production mechanisms: neutralino pair-production from on-shell 
$Z$-boson decays via the higgsino component of the neutralinos, and single neutralino production from $D$- or 
$B$-meson decays via an RPV $LQ\bar D$ coupling. In the study of neutralinos produced from a meson, we take two benchmark 
scenarios from  Refs.~\cite{deVries:2015mfw,Dercks:2018eua} for illustration of our results. Scenario 1 has the neutralino produced
from a $D_s$-meson decay while scenario 2 from a $B^0$-meson decay.

We present our HNL results in the mixing angle squared, $|V_{\alpha N}|^2$, vs. mass, $m_N$, plane, where $\alpha = e,\mu$,
\textit{cf.} Fig.~\ref{FIG:VsqMN}. We consider \texttt{AL3X} with 100/fb or 250/fb integrated luminosity, and compare with theoretical 
projections of other proposed detectors. We find that \texttt{AL3X} reaches smaller mixing angles than both \texttt{FASER} and 
\texttt{CODEX-b} in its whole mass reach, but is weaker than \texttt{MATHUSLA} by a factor $\sim 3$ for masses below $\sim 
4\,$GeV. Compared to 
\texttt{SHiP}, \texttt{AL3X} is worse in mixing angle reach by one order of magnitude for $m_N$ below the $D$-meson threshold,
$\sim 2\,$GeV, and is slightly better than \texttt{SHiP}, for larger mass values.

As for detecting neutralinos pair-produced from $Z$-boson decays, we present two plots respectively for BR$(Z\rightarrow 2\tilde{\chi}_1^0)=10^{-3}$ at the experimental upper limit and for BR$(Z\rightarrow 2\tilde{\chi}_1^0)=10^{-5}$, switching on a single 
$LQ\bar D$ coupling: $\lam'_{112}$, for the neutralino decay. The plots are shown in the plane $\nicefrac{\lam'_{112}}{m^2_
{\tilde{f}}}$ vs. $m_{\tilde{\chi}^0_1}$, \textit{cf.} Fig.~\ref{fig:zneuneuresults}. We find \texttt{AL3X}, comparable to other detectors, 
has a mass reach from $\sim 1$ GeV up to $\sim m_Z/2$. While \texttt{MATHUSLA} has the strongest reach in $\nicefrac{\lam'_{112}}
{m^2_{\tilde{f}}}$, \texttt{AL3X} is only slightly worse by a factor $\sim 2$. Novel parameter space, which is orders of magnitude 
more sensitive than the present experimental limits on $\nicefrac{\lam'_{112}}{m^2_{\tilde{f}}}$, can be probed by all of these detectors.

We show two sets of plots for the light neutralinos singly produced from a charm or bottom meson, where two $LQ\bar D$ couplings 
are switched on: $\lambda'_{\text{prod}}$ and $\lambda'_{\text{dec}}$, responsible for the production and the decay of the lightest 
neutralino, respectively. In the first set, Fig.~\ref{fig:rpv1}, shown  in the plane $\nicefrac{\lambda'_{\text{prod}}}{m^2_{\tilde{f}}}$ vs. $
\nicefrac{\lambda'_{\text{dec}}}{m^2_{\tilde{f}}}$ for three representative values of $m_{\tilde{\chi}^0_1}$, we find that in scenario 1, 
\texttt{AL3X} has a reach in $\nicefrac{\lambda'_{\text{prod}/\text{dec}}}{m^2_{\tilde{f}}}$ beyond existing bounds by roughly an 
order of magnitude, but weaker than \texttt{SHiP} \cite{deVries:2015mfw}, by approximately a factor of 3. In the other scenario 
associated with a $B^0$-meson, \texttt{AL3X} can go beyond existing limits by almost two orders of magnitude, and is more
sensitive than \texttt{SHiP} by about a factor of $5$ in both axes. In the second set of plots, Fig.~\ref{fig:rpv2}, we present results 
in the plane Br(meson$\,\rightarrow \tilde{\chi}^0_1$)$\cdot$Br($\tilde{\chi}^0_1\rightarrow$ visibles) vs. $c\tau$, the decay 
length of $\tilde{\chi}^0_1$, and compare with other experiments. In scenario 1, \texttt{SHiP} shows the strongest sensitivity in the 
product of branching ratios, covering all the sensitive areas of the other proposed detectors, while in scenario 2 \texttt{AL3X} 
supersedes the whole sensitivity region of \texttt{SHiP}, and complements \texttt{MATHUSLA} in different $c\tau$ regimes.

In summary, we conclude that \texttt{AL3X} can complement or even
exceed the other proposed detectors in the parameter space of
the different models considered here. It might be interesting to study
also other models for this newly proposed detector. Finally, we stress
that our sensitivity estimates are based on the assumption of
essentially background-free experimental searches.  Any unforeseen
background could seriously affect these conclusions.

\medskip

\textit{Note added:} While completing this work, an
  updated sensitivity estimate for \texttt{SHiP} has been published
  recently in Ref.~\cite{SHiP:2018xqw}.

%%%%%%%%%%%%%%%%%%%%%%%%%%%%%%%%%%%%%%%%%%%%%%%%%%%%%%%%%%%%%%%%%%%%%%
\bigskip
\centerline{\bf Acknowledgements}

\bigskip
We thank Torbj\"{o}rn Sj\"{o}strand, Jordy de Vries and Juan Carlos Helo for useful discussions.
H.K.D. and Z.S.W. are supported by the Sino-German DFG grant SFB CRC
110 ``Symmetries and the Emergence of Structure in QCD".  M.H. acknowledges
support by Spanish grants FPA2017-85216-P, SEV-2014-0398 (AEI/FEDER,
UE), Red Consolider MultiDark (FPA2017-090566-REDC) and
PROMETEOII/2018/165 (Generalitat Valenciana).

\bigskip
%%%%%%%%%%%%%%%%%%%%%%%%%%%%%%%%%%%%%%%%%%%%%%%%%%%%%%%%%%%%%%%%%%%%%%

\bibliography{Refs}
\bibliographystyle{h-physrev5}

\end{document}